\documentclass[letterpaper,12pt]{article}
\usepackage{osajnl}
\usepackage{amsmath,amssymb}

\renewcommand{\Re}{\textrm{Re}}
\renewcommand{\Im}{\textrm{Im}}

\begin{document}

\title{Approximate equivalence between guided modes in a low-contrast photonic bandgap fiber and Maxwell TM modes of a high-contrast 2D photonic structure}
\author{Olivier Legrand, Laurent Labont\'e, and Christian Vanneste}
\address{Laboratoire de Physique de la Mati\`ere Condens\'ee,
CNRS UMR 6622,\\  Universit\'e de Nice Sophia-Antipolis, 06108 Nice,
France}

\begin{abstract}
We present a formal analogy between the eigenvalue problem for guided scalar modes in a low contrast photonic bandgap fiber and quasi-stationary TM modes of a 2D photonic structure. Using this analogy, we numerically study the confinement losses of disordered microstructured fibers through the leakage rate of a open 2D system with high refractive index inclusions. Our results show that for large values of the disorder, the confinement losses increase. However, they also suggest that losses might be improved in strongly disordered fibers by exploring ranges of physical parameters where Anderson localization sets in.
\end{abstract}

\ocis{060.5295, 290.4210.}

\maketitle

\section{Introduction}
The equivalence between the guided modes of a microstructured fiber and the TM modes of a two-dimensional (2D) structure shows that a low index contrast in the fiber corresponds to a high contrast in the 2D structure. Moreover the guided modes of a microstructured fiber experience confinement losses due to transverse leakage. Hence, even in the absence of material absorption, the guided field decays slowly as a function of time. In order  to obtain a full equivalence between the lossy fiber and the 2D structure, the latter must also suffer leakage through the boundaries. In other words, the 2D structure must be open instead of being a closed cavity.  If the leakage through the boundaries is weak and can be considered as a small perturbation of the closed system, one can decompose the field over a basis of \emph{quasi-modes}, which are very close to the stationary modes of the closed 2D cavity, but slowly decay in time. Their  eigenvalues (or energies) are complex instead of being real  numbers for a closed system, the small imaginary part describing the exponential decay of the mode as a function of time.

Using this equivalence, we have studied the modes of an open disordered 2D structure in order to get the confinement loss of a microstructured fiber as a function of the position disorder of the high index inclusions. We observe that some realizations of the disorder display leakage rates smaller than the leakage rate of the periodic system in agreement with recent results by V. Pureur et al \cite{Pureur2007}. However, this noticeable effect is only observed for small amounts of disorder. When we increase the disorder at amounts larger than those studied previously, the average leaking rate eventually increases with the disorder. This result is compared with the predictions of Anderson localization theory for disordered systems.

\section{Equivalence between the modes of a fiber and TM modes of a 2D structure}
Let us first consider the calculation of quasi-stationary TM modes in an open 2D structure composed of circular scatterers with high refractive index $N_{scat}$ embedded in a host material of index unity (vacuum). Such a system is equivalent to an array of parallel cylinders with the electric field aligned with the axis of the cylinders. In the following, we consider a finite size triangular arrangement of rods from which the central rod is removed, corresponding to the geometry of commonly used low-contrast photonic bandgap fibers \cite{Argyros2005}. Due to the existence of Mie resonances, frequency gaps are likely to occur whose size and position are very little dependent on the arrangement (whether periodic or not) of the cylinders \cite{Soukoulis, Rockstuhl}. In the presence of a defect, as for instance removing a rod from the structure, sharp peaks are generally observed inside such bandgaps. These peaks are associated to quasi-modes which are spatially localized. Hence, the vacancy introduced by removing the central rod can be considered as a defect of the 2D structure. Therefore, one can expect the existence of long-lived quasi-modes located at the vacancy location, which are the 2D equivalent of the modes guided in the fiber core of the low-contrast photonic bandgap fiber.

\begin{figure}[htbp]
\begin{center}
\includegraphics[width=8.3cm]{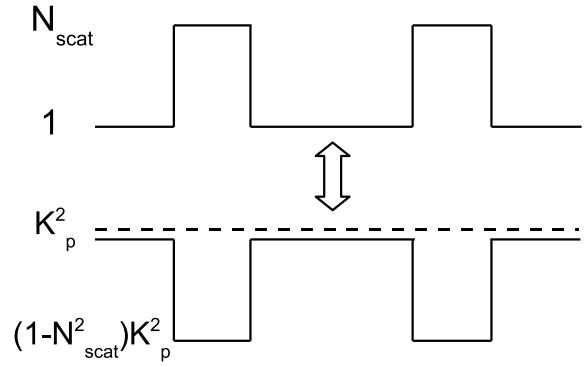}
\end{center}
\caption{Picture of the index profile and of the corresponding potential profile. Fig1.eps.}
\label{PotWell}
\end{figure}

At the frequency $\Omega_p = c K_p$ of a quasi-mode $p$, the component $\Psi$ of the electric field perpendicular to the plane of a 2D cavity can be written as $\Psi_p = \Phi_p (\vec{r}) \textrm{e}^{i\Omega_p t}$ satisfying the 2D Helmholtz equation 
\begin{equation}
\label{Helmholtz}
( \Delta + N^2(\vec{r})K_p^2 ) \Phi_p = 0
\end{equation}
It is worth noting that our 2D system is formally equivalent to the problem of a quantum particle with positive energy above a structured potential consisting of negative circular potential wells embedded in a zero potential.
This is easily seen by  re-writing Eq. (\ref{Helmholtz}) in the form of the stationary Schr\"odinger equation
\begin{equation}
\label{eigprob}
\left[-\Delta
+(1-N^2(\vec{r}))K_p^2\right]\,\Phi_p(\vec{r})
=K_p^2\,\Phi_p(\vec{r})\,.
\end{equation}
where $N(\vec{r})=1$ in the vacuum and $N(\vec{r})=N_{scat}$ inside the scatterers.
Therefore, the potential well of depth $V_{scat} = (1-N_{scat}^2)K_p^2$ associated to a dielectric scatterer should not be viewed as a confining well but rather as a resonant scattering potential well since $N_{scat} > 1$ and since the real part of the eigenvalue $K_p^2$ is positive (see Fig. \ref{PotWell}).

Let us now consider the guided modes in low-contrast photonic bandgap fibers consisting of high-index ($n_1$) rods in a low-index ($n_0$) cladding. At a given frequency $\nu$ defining a vacuum wavelength $\lambda = c /\nu$ and the associated wavenumber $k = 2 \pi / \lambda$, propagating (scalar) quasi-modes can be written as 
$\psi_p = \phi_p (\vec{r}_{\perp}) \exp [i k n^{eff}_p z]$
satisfying the Helmholtz equation 
\begin{equation}
\label{Helmholtz2}
( \Delta_{\perp} -k^2 (n^{eff}_p)^2 ) \phi_p + n^2 (\vec{r}_{\perp}) k^2 \phi_p= 0
\end{equation}
which can be re-written
\begin{equation}
\label{eigprob2}
- \Delta_{\perp}  \phi_p + [n_0^2 - n^2 (\vec{r}_{\perp}) ] k^2 \phi_p = [ n_0^2 - (n^{eff}_p)^2 ] k^2 \phi_p 
\end{equation}
where, in the present context, $n (\vec{r}_{\perp}) = n_0$ in the cladding and  $n (\vec{r}_{\perp}) = n_1$ inside the rods. This way of writing the Helmholtz equation emphasizes the analogy with the quantum problem introduced above where the cladding index defines the \emph{zero} potential. Note that we are considering quasi-modes of the ARROW type \cite{Argyros2005} (corresponding to the resonant scattering effect mentioned above) implying $n^{eff}_p < n_0$ as opposed to the \emph{LP-modes} guided inside the high-index rods $n^{eff}_p > n_0$ \cite{Snyder, Marcuse}. Moreover, in this description, the scalar hypothesis is justified in the weak guidance limit (see e.g. \cite{Snyder&Love}).
Putting together the latter conditions, one obtains 
\begin{equation}
0 < n_0^2 - (n^{eff}_p)^2 \ll  1 \, .
\end{equation}

When one compares equations (\ref{eigprob}) and (\ref{eigprob2}), one clearly sees that a solution $\Phi_p$ of (\ref{eigprob}) for a given eigenvalue $K_p^2$ can be considered as a solution of (\ref{eigprob2}) if one identifies 
\begin{subequations}
\begin{eqnarray}
[n_0^2 - (n^{eff}_p)^2]  k^2 &\textrm{with}& K_p^2 \quad \textrm{and} \\
(n_0^2 - n_1^2 ) k^2 &\textrm{with}& (1-N_{scat}^2)K_p^2 \, , 
\end{eqnarray}
\end{subequations}
together with the condition $ n_1^2 - n_0^2 \ll (N_{scat}^2 - 1)$ according to the weak guidance limit for which $K_p^2 \ll k^2$. The latter condition indicates that TM solutions of the 2D problem can be mapped onto solutions in the photonic bandgap fiber only if $N_{scat}^2$ is much larger than unity. The principal difference between both problems resides in the fact that the unknown eigenvalue $K_p^2$ is factorized in the potential term $(1-N^2(\vec{r}))K_p^2$ of equation (\ref{eigprob}) whereas the unknown effective index $n^{eff}_p$ does not appear in the potential term $[n_0^2 - n^2 (\vec{r}_{\perp}) ] k^2$ of equation (\ref{eigprob2}). Note that the above equivalence assumes that the geometries of both systems are identical.

Within the bandgaps, one can also establish a correspondence between the loss along the fiber and the decay rate of the $2D$ problem. This is readily done by expanding the supposedly small imaginary parts of  $[n_0^2 - (n^{eff}_p)^2]  k^2$ and $K_p^2$ and by identification, one gets:
\begin{equation}
k^2 \, \Re{\, n^{eff}}\, \Im{\, n^{eff}} = \Re{K} \, \Im{K} = \frac{\Omega}{2 c^2} \Gamma
\end{equation}
where $\Gamma / 2$ is the amplitude decay rate of the $2D$ quasi-stationary TM mode. 

\section{Numerical study}
We have numerically studied the 2D system with the finite-difference time-domain method (FDTD). Open boundary conditions are modeled by perfectly matched layers (PML). We first investigated the unperturbed system where the scatterers are positioned over a periodic triangular lattice (Fig. 2).

Calculations proceed in two steps. In the first step, one obtains the impulse response by exciting the system with a wide-band pulse. By Fourier transform of the recorded field over successive time windows, one observes the evolution of the spectrum as a function of the time. A large amount of the initial energy escapes rapidly from the open system. Only the small leakage resonances persist at long times. In the second step, one excites the same system by a monochromatic source at the frequencies of the narrowest peaks in the impulse spectrum. In the following, we will present results for the two narrowest modes hereafter denoted modes 1 and 2. These two modes are localized at the center of the system (Fig.~2). After the build-up of the corresponding resonance, the source is stopped. Since the system is open, the energy starts decreasing as a function of time. By measuring the decay of the field amplitude, one obtains the lifetime of the resonance. Using the mapping described above, one obtains the leakage rate of the corresponding fiber mode.

Next, we resume the same calculations after introducing some amount of disorder on the location of the scatterers by perturbation of the periodic array. A random displacement with adjustable amplitude $\delta x$ and $\delta y$ is imposed to each scatterer $i$ of the system. The amount of disorder relative to the periodic system is given by the mean displacement $\sigma $ over the whole set of scatterers where $\sigma^2 = \frac{1}{N}\sum_{i=1}^{N}( \delta x_i^2 + \delta y_i^2 )$. For each value of this parameter, several implementations of the disorder have been investigated.

\section{Results and Conclusion}
The 2D system is a periodic array of circular scatterers whose refractive index is $N_{scat}=4$ in a background medium of refractive index $N=1$. 
The array pitch is 15 $\mu$m and the radius of the scatterers is $r=5$ $\mu$m. The system is made of seven layers of scatterers (Fig. 2). The amount of disorder on the positions of the scatterers ranges from $\sigma=0$ to $\sigma=3.5$ $\mu$m. Note that the latter value corresponds to a noticeable perturbation of the system as shown in Fig. 2. Note also that the above values of the refractive index in the $2D$ system would correspond, for instance, to $n_0 = 1.450$ and $n_1 = 1.467$ at $\lambda = 1.6$ $\mu$m for mode 1 in the fiber problem.

Figure 3 displays the attenuation rate of modes 1 (lower curves) and 2 (upper curves) as a function of the amount of disorder for each of the different systems, which have been investigated. The lines connecting the data points enable us to visualize the modes 1 and 2, which correspond to the same realizations of the disorder. The values that have been obtained for the periodic system ($\sigma=0$) are in good agreement with the values obtained for the same system by an independent calculation using the finite-element method.

We note that large fluctuations take place between different realizations of the disorder. For $\sigma$ values that do not exceed about $1.5 \mu$m, some fluctuations correspond to attenuation rates smaller than the values of the periodic system. This is a noticeable result, which agrees with those of Pureur et al \cite{Pureur2007}. Hence, against naive intuition, a slightly disordered system can display a better confinement than the corresponding unperturbed system. However, Fig. 3 shows that for large amounts of disorder the mean leakage rate eventually increases as expected.

It is interesting to point out that an anticorrelation is observed between the losses of modes 1 and 2 for the smallest values of the disorder parameter ($\sigma<1.5\mu $m). When the losses of mode 1 vary with $\sigma$, the losses of mode 2 vary systematically in the opposite way for the corresponding realizations of the disorder as highlighted by the lines joining the data points in Fig. 3. 

The fact that the average losses increase with $\sigma $ rules out the possiblity for the system to be in the localized regime due to disorder. Indeed, in this regime, modes are characterized by a typical size, the localization length, which is smaller than the system size. The shorter the localization length is, i.e. the better a localized mode is confined at the center of the system, the smaller the decay rate due to a weaker leakage through the boundaries \cite{Laurent2007}. As the localization length is expected to decrease when the disorder increases, well localized modes should exhibit attenuation rates, which decrease with the amount of disorder contrary to the behavior observed in Fig. 3. Note that for weak disorder, leakage rates smaller than that of the periodic system correspond to fluctuations due to particular realizations of the disorder. They do not correspond to a general trend of decreasing losses with increasing disorder.

Obviously, these results depend on the parameters of the systems we have considered. One can wonder whether the localization regime, which would improve the confinement losses of a microstructured fiber is attainable by varying the system parameters. For instance, it is well known that sufficiently large 2D disordered systems are always in the localization regime \cite{Scaling1979}. Hence, by increasing the number of layers of scatterers, the system size will eventually exceed the localization length and result in reduced losses. However, such an achievement seems to be out of reach of the current possibilities of fabrication of microstructured fibers. Instead of varying the system size, one could change the refractive index of the scatterers. 
Figure 4 displays the localization length as a function of the $V$-parameter for scatterers of radius $5$ $\mu$m. The $V$-parameter is commonly used in the context of fibers and is defined by
\begin{equation}
V = \frac{2 \pi r (n_1^2 - n_0^2 )^{1/2}}{\lambda}
\end{equation}
where $r$ is the radius of the scatterers (see e.g. \cite{Argyros2005}). The different curves in Fig. 4 correspond to different values of the refractive index of the scatterers $N_{scat}=3, 4, 5, 6$. For values of the $V$-parameter close to $V=4$, which correspond to mode 1 of the fibers studied in this article, the localization length is very sensitive to the value of the refractive index, ranging from $\xi = 10^{5}$ $\mu$m for $N_{scat}=3$ to $\xi = 10^{2}$ $\mu$m for $N_{scat}=6$. The latter value of $\xi$, which is smaller than the fiber diameter $D \simeq 200$ $\mu$m for the system shown in Fig. 2, indicates that localization might be reachable for high values of the refractive indices of the scatterers.

The above estimates of the localization length have been obtained from the scattering cross section of 2D Mie scatterers \cite{VandeHulst} in the independent scattering approximation \cite{Derode2001}. This last approximation is certainly not well verified for the values of the pitch and of  the scatterer diameters considered here. Hence, the above localization lengths are only indicative. However, the sensitivity of the localization length to the index contrast suggests that there is some hope to reduce the confinement losses of disordered microstructured fibers by carefully exploring the ranges of their physical parameters.

We acknowledge support from the project ANR-05-BLAN-0080 ``FOCALASE".

\clearpage

\section*{List of Figure Captions}

Fig. 1. Picture of the index profile and of the corresponding potential profile.

\noindent Fig. 2. (Color online) Periodic system and an example of a disordered system: the map of a mode is shown together with its amplitude (Log scale) along a section marked on the map.

\noindent Fig. 3. Attenuation rates of fiber modes 1 and 2 as a function of the amount of disorder $\sigma$.

\noindent Fig. 4. The localization length as a function of the $V$-parameter for scatterers of radius $5 \mu$m. The different curves, labeled by $n=3.0, 4.0, 5.0, 6.0$, correspond to different values of the refractive index of the scatterers.

\listoffigures

\clearpage

\clearpage

\begin{figure}[htbp]
\begin{center}
\includegraphics[width=8.3cm]{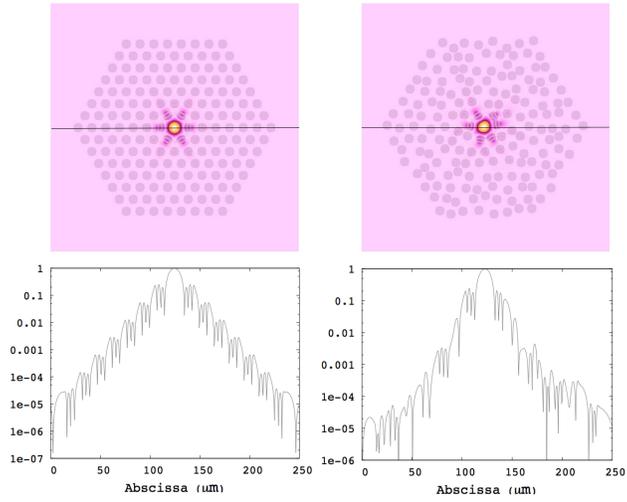}
\caption{(Color online) Periodic system and an example of a disordered system: the map of a mode is shown together with its amplitude (logarithmic scale) along a section marked on the map. Fig2.eps.}
\end{center}
\end{figure}

\clearpage

\begin{figure}[htbp]
\begin{center}
\includegraphics[width=8.3cm]{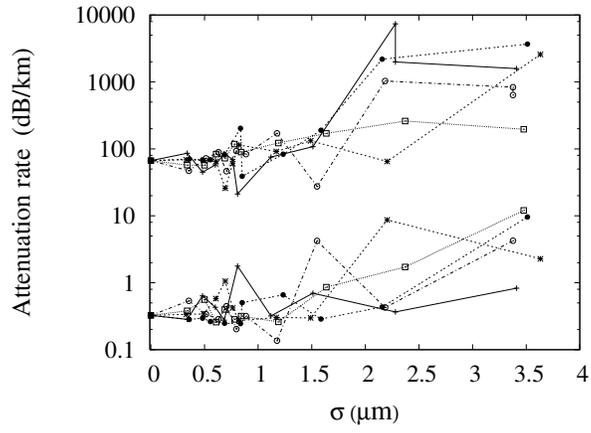}
\caption{Attenuation rates of fiber modes 1 and 2 as a function of the amount of disorder $\sigma$. Fig3.eps.}
\end{center}
\end{figure}

\clearpage

\begin{figure}[htbp]
\begin{center}
\includegraphics[width=8.3cm]{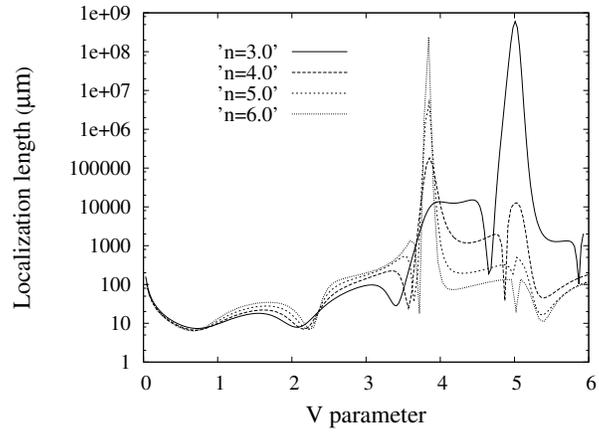}
\caption{ The localization length as a function of the $V$-parameter for scatterers of radius $5$ $\mu$m. The different curves, labeled by $n=3.0, 4.0, 5.0, 6.0$, correspond to different values of the refractive index of the scatterers. Fig4.eps.}
\end{center}
\end{figure}

\end{document}